\documentclass[aps,pre,twocolumn,showpacs,floatfix,preprintnumbers,amsmath,amssymb]{revtex4-1}
\usepackage{amsmath}
\usepackage{amssymb}
\usepackage{graphicx}
\usepackage{epstopdf}
\usepackage{epsfig}
\usepackage{dcolumn}
\usepackage{bm}

\begin{document}
\title{Floquet states of kicked particle in a singular potential: Exponential and power-law profiles}
\author{Sanku Paul}
\email{Email: sankup005@gmail.com}
\affiliation{
Indian Institute of Science Education and Research, Dr. Homi Bhabha Road, Pune 411 008, India.}
\author{M. S. Santhanam}
\email{Email: santh@iiserpune.ac.in}
\affiliation{
Indian Institute of Science Education and Research, Dr. Homi Bhabha Road, Pune 411 008, India.}

\date{\today}
\begin{abstract}
It is well known that, in the chaotic regime, all the Floquet states of kicked rotor 
system display an exponential profile resulting from dynamical localization. 
If the kicked rotor is placed in an additional stationary infinite potential well,
its Floquet states display power-law profile.
It has also been suggested in general that the Floquet states of periodically kicked systems with
singularities in the potential would have power-law profile.
In this work, we study the Floquet states of a kicked particle
in {\it finite} potential barrier. By varying the height of finite potential
barrier, the nature of transition in the Floquet
state from exponential to power-law decay profile is studied. 
We map this system to a tight binding model
and show that the nature of decay profile depends on
energy band spanned by the Floquet states (in unperturbed basis)
relative to the potential height. 
This property can also be inferred from the
statistics of Floquet eigenvalues and eigenvectors. This leads to an unusual scenario in which the
level spacing distribution, as a window in to the spectral correlations,
is not a unique characteristic for the entire system.
\end{abstract}

\maketitle

\section{Introduction}

Dynamical localization of Floquet states in time-dependent and chaotic Hamiltonian systems
is a phase coherent effect arising from quantum interferences. Quantum kicked rotor 
is a paradigmatic model for quantized chaotic systems which displays localization effects.
Quantum localization in kicked rotor (KR) continues to attract attention
in a variety of contexts ranging from metal-insulator transitions \cite{chabe, garcia}, coherent control \cite{gong},
entanglement measures \cite{santh}, quantum resonances \cite{fish1, max, beh, dana1}, quantum ratchets \cite{schanz, dana} to quantum transport \cite{jones}
and decoherence effects \cite{sarkar, lutz}. Most of such studies have focussed on KR as a model for time-dependent potential exhibiting classically chaotic
dynamics and quantum mechanical localization. For sufficiently strong non-linearity,
KR displays chaotic classical dynamics and it is associated with diffusive growth of mean energy
with time. In the corresponding quantum regime, this unbounded energy growth
is strongly suppressed by localization arising from destructive quantum interferences \cite{kr}.
This effect in KR has been shown to be analogous to Anderson localization for electronic
transport in crystalline solids \cite{fishman, stockmann, reichl}.

One significant property shared by the quantum KR and Anderson model is
the exponential decay of their eigenstates. In the one dimensional Anderson model all
the eigenstates are exponentially localized in position representation \cite{anderson1, anderson2}, 
i.e, $\psi(x) \sim e^{-x/x_l}$ where $x_l$ is the localization length. In the KR
system, eigenstates are exponentially localized in the momentum representation \cite{dima}.
The latter has been experimentally realized in microwave ionization of hydrogen atoms and 
in cold atomic cloud in optical lattices \cite{moore, delande, fromhold}.

Kicked rotor can thus be regarded as a representative dynamical system from two distinct points of view.
Firstly, in the classical sense, it belongs to a class of chaotic systems that obeys 
Kolmogorov-Arnold-Moser (KAM) theorem \cite{jurgen}. This effectively implies that, upon
variation of a chaos parameter, the system makes a smooth transition from regular to
predominantly chaotic dynamics. Secondly, in the quantum mechanical regime,
KR is a paradigmatic example of dynamical localization and the associated exponential
profile of its Floquet states. In the last one decade, many other facets of chaos and localization
in variants of KR have been studied that have provided results different
from this standard scenario \cite{bala, italo, pragya, klappauf, garcia}.

One class of important variant is to place the KR in a singular
potential. Presence of singularity in the potential violates one of the conditions
for the applicability of KAM theorem and leads to a scenario in which abrupt,
rather than smooth, transition
from integrability to chaotic dynamics becomes possible. Such abrupt transition to chaos
is a feature of non-KAM systems and is seen, for instance, in the
kicked particle in an infinite potential well \cite{sankar, hu}. The quantum eigenstates of this
system had been reported to display localization and its profile is {\sl not}
exponential but was claimed to have power-law type decay in the unperturbed basis.
A more systematic study in Ref. \cite{garcia} incorporated singularity in the KR
through a tunable potential term $V(q;\alpha)$ such that it becomes singular at some special
value of tunable parameter $\alpha= \alpha_s$. It was shown, through numerical simulations,
that if $\alpha=\alpha_s$ in the potential, then all the eigenstates of the system are
power-law localized. Indeed, it was even suggested that KR when acted upon by
a singular potential would display eigenstate localization with power-law profile
in contrast to the exponential profile obtained in the context of standard KR \cite{hu, liu}.
This suggestion has not yet been numerically tested in a variety of chaotic
Hamiltonian systems and general analytical results in support for this claim remains
an open question.

In this paper, we examine the question whether the presence of non-analytic potential
in a kicked rotor would {\it generically} imply power-law profile for its eigenstates
in the quantum regime. To address this question, we consider the dynamics of a periodically kicked particle
placed in a stationary finite potential well of height $V_0$. This is primarily a non-KAM system
and its unusual classical and quantum transport properties, reflective of its
non-KAM nature, were recently reported in Ref. \cite{paul}. This system subsumes two limiting cases; it
is the standard KR (a KAM system) in the absence of finite well potential, i.e, $V_0 = 0$ and
if $V_0 \to \infty$, then it becomes a kicked rotor system placed in an infinite well
(a non-KAM system) and has been studied in Refs. \cite{sankar, hu}.
Hence, this is a suitable test bed to understand the transition in the nature of
Floquet states as $V_0$ is varied from the
limit of a KR system (analytic potential) to that of a system with 
singular potential. Further, this can lead to a better understanding of the quantum manifestations  of
classical chaos non-KAM systems.

\begin{figure}
\includegraphics*[width=3.0in]{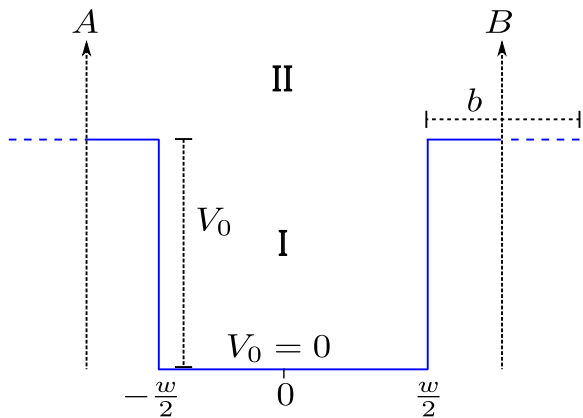}
\caption{(Color online) Schematic of the stationary
potential, $V_{sq}(\theta)$ with $V_0$ as the potential height, $b$ and $w$ as barrier and well width. $A$ and $B$ represents the positions at which periodic boundary conditions are applied. I and II denotes the regions below and above $V_0$.}
\label{fig1}
\end{figure}

Using the context of this system based on KR, we show in this paper that the
presence of singularity in the potential does not always guarantee
power-law localization of Floquet states.
Singular potentials are associated with power-law localized Floquet states provided
that the Floquet states span an energy band in which the singularity is effectively
felt by the particle. Further, it is demonstrated that the spectral
fluctuations properties such as the level spacing distributions for this
system depends on the energy range being considered. Hence, spacing distributions
do not characterize the system at all the energy scales.

In section 2, we introduce the model of kicked particle in a finite barrier and in
section 3 we report results on the decay profile of the Floquet states and relate
it to the decay of the Floquet matrix and to the effective singularity ``felt" by the
kicked particle at various energy scales. In section 4, we obtain a tight binding
form for our system to deduce the non-exponential nature of Floquet state decays.
Finally, in section 5, we discuss the manifestation of potential singularity in 
the averaged quantities derived from Floquet states.

\section{Kicked particle in finite barrier}

The dimensionless Hamiltonian of a periodically kicked particle in a finite well
potential \cite{paul} is
\begin{equation}
\begin{aligned}
H &= \frac{p^2}{2}+V_{sq}(\theta)+k~\cos(\theta) \sum_{n=-\infty}^{\infty} \delta(t-n)\\
 &= H_0 + V(\theta) \sum_{n=-\infty}^{\infty} \delta(t-n).
\end{aligned}
\label{ham1}
\end{equation}
In this, $V(\theta)=k~\cos(\theta)$ and $V_{sq}(\theta)$ is the square well potential shown in Fig. \ref{fig1} and can
be represented as
\begin{equation*}
V_{sq}(\theta)=V_0[\Theta(\theta-R\pi)-\Theta(\theta-R\pi-b)],
\end{equation*}
where $V_0$ is the potential height and $b$ is the barrier width,
$R=w/\lambda$ is the ratio of well width to the wavelength of the kicking field and $k$ is the kick 
strength. Throughout this work, we have set $\lambda=2\pi$, $b$ and $w$ are constrained by
$b+w=2\pi$. Periodic boundary conditions are applied at positions A and B shown in Fig. \ref{fig1}.

Let $E_n$ and $|\psi_n\rangle$ represent the energy and the eigenstate of
the unperturbed system such that $H_0|\psi_n \rangle = E_n |\psi_n \rangle$.
Further, $|\psi_n\rangle$ can be written as a superposition of all momentum
states $|l\rangle$, i.e. $|\psi_n \rangle=\sum_l a_{nl}|l\rangle$, where $a_{nl}$ represents the 
expansion coefficient. Then any general initial state can be expressed
in the energy basis state representation as $|\Psi\rangle=\sum_n b_n|\psi_n\rangle$.
The mean energy in the state $|\Psi(t)\rangle$ can be obtained as

 \begin{equation}
 \begin{aligned}
 E(t) &= \langle\Psi(t)|\hat{H_0}|\Psi(t)\rangle \\
  &=  \sum_m E_m |b_m(t)|^2.
 \end{aligned}
 \end{equation}
The quantum map that connects the state $|\Psi(N+1)\rangle$ at time $N+1$ with the 
state $|\Psi(N)\rangle$ can be obtained by evolving the Schroedinger equation
$|\Psi(N+1)\rangle=\widehat{U}|\Psi(N)\rangle$,
where $\widehat{U}$ is the Floquet operator, 
\begin{equation}
\widehat{U}=e^{-H_0/\hbar_s}e^{-iV(\theta)/\hbar_s},
\end{equation}
and $\hbar_s$ is the scaled Planck's constant. In the energy representation,
the elements of the Floquet operator are given by
\begin{equation}
U_{nm}=\sum_{p,p'} a_{np}^{*} a_{mp'} i^{|p-p'|}J_{|p-p'|}\left(\frac{k}{\hbar_s}\right),
\label{umat}
\end{equation}
where $J_{|p-p'|}(\cdot)$ is the Bessel function of order $|p-p'|$.

The eigenvalue equation governing the Floquet operator is $\widehat{U} |\phi\rangle = e^{i \omega} |\phi\rangle$
in which $|\phi\rangle$ represents a Floquet state and $\omega$ is its quasi-energy. The 
Floquet operator is an unitary operator and hence the eigenvalues lie
on a unit circle. Further the quasi-energy state, $|\phi\rangle$, can be decomposed as a 
superposition of all energy states $|\psi_n\rangle$, i.e., 
$|\phi\rangle = \sum_n c_n |\psi_n\rangle$, where $|c_n|$ is the
probability density of finding the particle in state $|\psi_n\rangle$.

In order to analyse the localization properties of the
Floquet states, Floquet matrix of order $N$ is numerically diagonalized to determine  the
quasi-energies and the Floquet vectors. In this work, $N=10035$ and we have ensured 
that for the choice of parameters used in this paper, the system is classically chaotic \cite{supplement}.
Floquet states for standard KR are generally known to be exponentially localized in
momentum space $|\psi(p)|^2 \sim \exp(-p/\xi)$ characterized by a localization length $\xi$. In
contrast to that, Floquet states of a kicked particle in a periodic potential well is localized over
energy basis state $|\psi_n\rangle, n=1,2,\dots$. In the subsequent sections,
it is shown that the system in Eq. \ref{ham1} exhibits a transition from exponential
to power-law localization as the parameters $V_0$ and $k$ are varied.

\section{Floquet states}

\begin{figure}
\centering
\includegraphics*[width=3.3in]{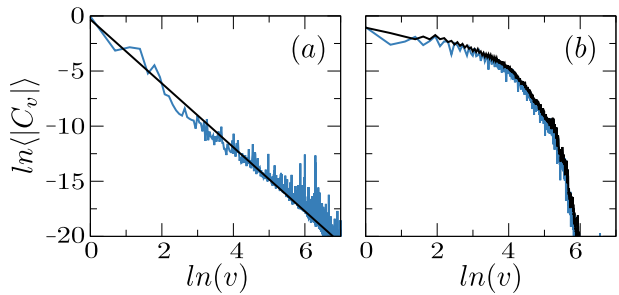}
\caption{(Color online) Decay of Floquet states over the unperturbed basis states, averaged
over all the Floquet states. The parameters are $b=1.4\pi, \hbar_s = 1.0$, (a) (KRIW limit) 
$V_0=5000.0, k=0.25$, and (b) (KR limit) $V_0=0.5, k=4.25$. In (a) black line is a linear fit and in (b) black curve corresponds to KR.}
\label{avgdFloquet states}
\end{figure}

In this section, we will mainly focus on the average spectral properties of the
Floquet states which governs the dynamics in the quantum regime.
For the finite well represented by Hamiltonian in Eq. \ref{ham1}, the nature of
Floquet state decay profile, in general, will depend on the choice of parameters, namely,
kick strength $k$ and potential height $V_0$.
Fig. \ref{avgdFloquet states} has been obtained by averaging over $10035$
Floquet states $|\phi\rangle(=\sum_n c_n |\psi_n\rangle)$ for each set of
parameters. Prior to averaging, each Floquet state was shifted by $n_{max}$, i.e. $v = n-n_{max}$, where $n_{max}$
corresponds to $n$ for which $|c_n|^2$ is maximum.

If $V_0 > 0$, the Hamiltonian in Eq. \ref{ham1} is a non-KAM system due to the presence of
singularities in $V_{sq}(\theta)$. Based on numerical simulations of kicked systems with singular
potentials, it was argued that their Floquet states display power-law decay over the
unperturbed basis \cite{sankar, hu, liu, garcia, jose}. Further, a new universality class has been proposed in Ref. \cite{garcia} based on the
presence of classical singularity and power-law localization.
To discuss the results, in the light of this proposal, two limiting cases can be identified;
(i) $0< V_0 < 1$ (KR limit) and
(ii) $V_0 \gg 1$ (KR in infinite well (KRIW) limit).
In the KR limit, notwithstanding the singularity in the potential, the Floquet states can be 
expected to be qualitatively closer to that of KR.
In particular, if kick strength $k \gg 1$, all the Floquet states display exponential decay profile.
On the other hand, in the KRIW limit, the potential height is large ($V_0 \gg 1$) and is 
qualitatively closer to the kicked infinite well system \cite{sankar, hu}. In this limit, even for small kick strengths 
$k<1$, it is known that all the Floquet states show power-law decay over the 
unperturbed basis \cite{hu, liu}. Both these limits are illustrated in Fig. \ref{avgdFloquet states}.

In Fig. \ref{avgdFloquet states}(a), the decay of the averaged Floquet state in the KRIW limit is
shown for $V_0=5000.0$ and $k=0.25$. It is consistent with a power-law form $P(v) \sim v^{-\gamma}$,
where $v > 0$ and $\gamma \approx 2.5$, in agreement with the value reported in Ref. \cite{hu} and the deviation observed can be attributed to the finite height of well.
On the other hand, averaged Floquet state in the KR limit for $V_0 = 0.5$
and $k = 0.25$ shown in Fig. \ref{avgdFloquet states}(b) displays exponential decay, $P(v) \sim \exp(-v/l)$, where $l$ is the
localization length. This is the standard dynamical localization scenario but is generally
not associated with non-KAM systems.
Both these decay profiles in Fig. \ref{avgdFloquet states} can be understood if the relation
between singular potential and power-law localization can be restated in the following
manner. For this purpose, let $\epsilon_{if}= \{E_i, E_{i+1}, E_{i+2}, \dots E_f\} $ 
collectively represent the energies of a set of states of $H_0$ lying in the energy band $(E_f-E_i)$
between two states with quantum numbers $f$ and $i$
in the unperturbed system. The classical singularities are associated with quantum
power-law localization of a set of Floquet states mostly lying in the energy range $\epsilon_{if}$,
provided $\epsilon_{if} < V_0$. Thus, Floquet states will display power-law localization
only if they effectively ``feel" the non-smooth potential. This requires that
energy scale $\epsilon_{if}$ be less than that representing $V_0$.

It must be emphasised that the Hamiltonian in Eq. \ref{ham1}
is classically a non-KAM system if $V_0 > 0$, for all values of $k>0$.
Hence, with $V_0=5000.0$ and for kick strength as small
as $k=0.25$ the system is classically chaotic \cite{supplement} and the corresponding quantized
system displays power-law localized profile of the Floquet states (Fig. \ref{avgdFloquet states}(a)).
In this case, most of the $10035$ Floquet states used for averaging are such that $\epsilon_{if} < V_0$.
On the other hand, in the
case of Fig. \ref{avgdFloquet states}(b), even though it is still a non-KAM system with
singular potential, the Floquet states mostly straddle energy scales $\epsilon_{if}$ larger than
$V_0$ and are not affected by the shallow singular potential and hence
the exponentially localized Floquet states are obtained.

\begin{figure}
\centering
\includegraphics*[width=3.3in]{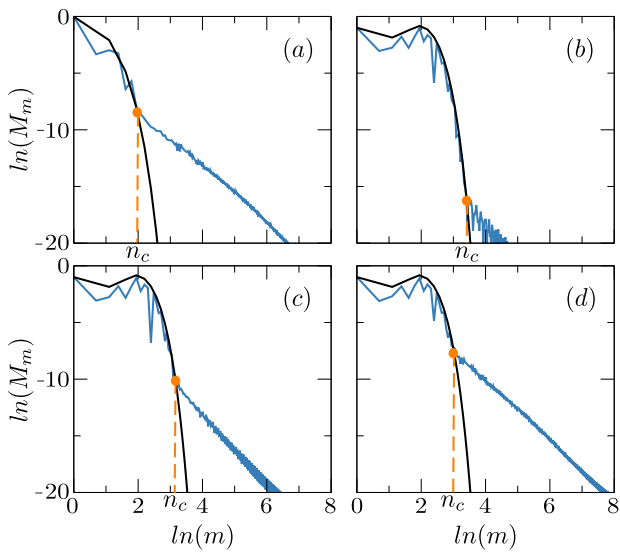}
\caption{(Color online) Averaged decay of the matrix elements of the Floquet operator $\widehat{U}$
as a function of $m$. The parameters are $b=1.4\pi, \hbar_s = 1.0$, (a) (KRIW limit) 
$V_0=5000.0, k=0.25$, and (b) (KR limit) $V_0=0.5, k=4.25$, (c) $V_0=100.0, k=4.25$,
(d) $V_0=5000.0, k=4.25$. $n_c$ represents the crossover point from exponential to power-law profile. All black curves corresponds to KR.}
\label{mele}
\end{figure}

\subsection{Matrix element decay}
It is known that exponential localization in the KR is associated with
the exponential decay of the matrix elements of the corresponding Floquet operator.
Further, from Ref. \cite{sankar, hu}, it is also known that in the case of KR in the
infinite well, a non-KAM system, the matrix elements of $\widehat{U}$ display a 
power-law decay after a bandwidth $\eta\propto k$.
Hence, it is natural to enquire how the decay of matrix elements changes
its character as $V_0>>1$ approaches the limit $V_0 \to 0$. In the unperturbed basis, the
matrix elements are $U_{nm}$ as given by Eq. \ref{umat}.
This is illustrated in Fig. \ref{mele} which shows $M_m = \langle |U_{nm}| \rangle_n$
as a function of $m$, with $m>n$, in log-log plot.

Figure \ref{mele}(a,b) shows $M_n$ as log-log plot for the same choice of parameters as in 
Fig. \ref{avgdFloquet states}(a,b). Figure \ref{mele}(a) corresponds to KRIW limit and
shows a short regime of exponential decay followed by an asymptotic power-law decay.
In Fig. \ref{mele}(b), $V_0=0.5$ corresponding to the KR limit and the decay of $M_n$
largely follows that of KR except for $n>>1$ where it decays as a power-law.
In general, the following features are observed. In the limit as $V_0 \to \infty$,
the decay is of power-law form. In the opposite limit of $V_0 \to 0$, the decay
is exponential in nature. In general, for any intermediate $V_0$, i.e., $0 < V_0 < \infty$, 
an initial exponential decay is followed by an asymptotic power-law decay whose
slope is approximately 2.7.
If $V_0 < \infty$, the initial exponential decay is always present. 
The exponential decay sharply changes over
to a power-law decay at $n=n_c$ as shown by dotted vertical lines in Fig. \ref{mele}.
For any fixed value of kick strength $k$, as $V_0$ varies from 0 $\to \infty$, then $n_c$
changes from $\infty  \to 0$. It is also to be noted that for fixed $V_0$, as $k$
increases, $n_c$ also increases.

\begin{figure*}[t]
\centering
\includegraphics*[width=6.2 in]{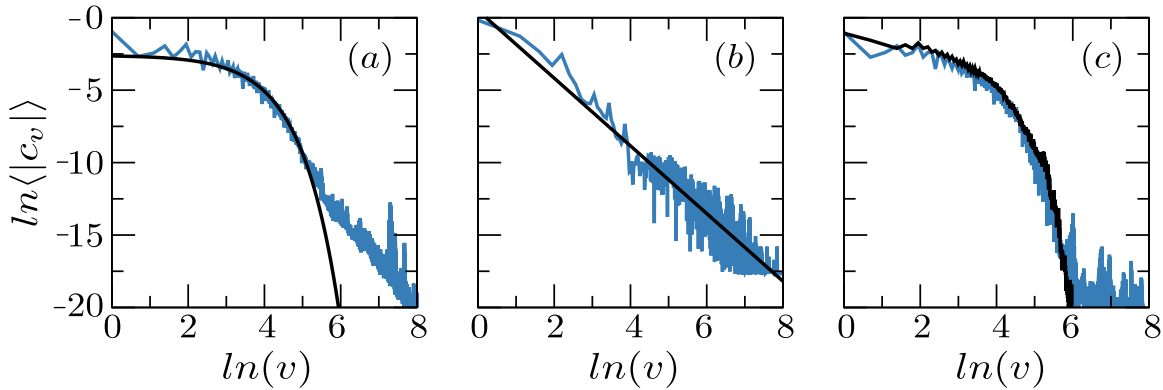}
\caption{(Color online) Floquet states of Hamiltonian in Eq. \ref{ham1} with parameters $b=1.4\pi$, $\hbar_s=1.0$, 
$V_0=5000.0$, $k=4.25$. The three figures differ in how Floquet states were averaged over;
(a) averaged over all the computed states, (b) averaged over states with $\mu< 1$, 
(c) averaged over states with $\mu \gg 1$. In (a,b) black curve represent best fit line and in (c) black curve corresponds to KR.}
\label{avgdrm}
\end{figure*}

\subsection{Energy scales}

In this section, we show how singularity of the potential and energy scales associated
with the Floquet states determine the localization structure of these states.
As discussed in Sec. 2, $c_{rj} = \langle \phi_r | \psi_j \rangle$,
where $|\psi_j\rangle$ is an eigenstate of $H_0$ with
associated energy $E_j$. Let $c_{max} = \mbox{max}(|c_{r,1}|^2, |c_{r,2}|^2, \dots |c_{r,N}|^2)$
represent the largest overlap of $r$-th Floquet state with $|\psi_j\rangle$. The energy
associated with $|\psi_j\rangle$, and hence with $c_{max}$, is denoted by $E_{max}$.
Then, an effective parameter $\mu=E_{max}/V_0$ can be identified to distinguish two regimes,
namely, (i) $\mu \le 1$ and (ii) $\mu >> 1$. Physically, $\mu \le 1$ corresponds to
Floquet states (in energy basis) mostly confined to the potential height $V_0$ and $\mu >>1$
corresponds to those Floquet states that have significant overlap with states lying
in the energy scales far greater than $V_0$.

In Fig. \ref{avgdrm}(a), $\ln \langle c_{v} \rangle$ is shown by averaging
over all the computed Floquet states for $V_0=5000.0$ and $k=4.25$. As discussed
earlier, the Floquet state profile is a combination of initial exponential decay
followed by a power-law decay. However, if average is taken only over those states
that satisfy the condition $\mu \le 1$, then the resulting profile is shown
in Fig. \ref{avgdrm}(b).
In this case $k=4.25$ and all the Floquet states are confined
to an energy scale well below $V_0$. Hence, these set of Floquet states can be
expected to ``feel" the presence of singularity in the potential. In this
regime, we observe a power-law profile for the averaged Floquet states as displayed in Fig. \ref{avgdrm}(b).
 
However if the states are averaged subject to the condition that $\mu >> 1$, then
singularity is not strongly felt by the Floquet states since the bandwidth of their energy
distribution is far greater than $V_0$. Effectively, at this
energy scale, the singularity becomes insignificant and hence we can expect it to 
lie in the KR limit. Indeed, as seen in Fig. \ref{avgdrm}(c), 
$\langle c_{v} \rangle$ is nearly identical to that of KR (shown as black curve in Fig. \ref{avgdrm}(c)) at $k=4.25$.

In general, for the Hamiltonian in Eq. \ref{ham1}, the localization property of a subset 
of Floquet states in an energy band $\epsilon_{if}$ depends on the effectiveness of 
the singularity for the spectral range $\epsilon_{if}$ under consideration.
In a given energy band $\epsilon_{if}$, if the singularity is effective, then 
power-law localization is obtained and if singularity is weak or absent then 
exponential localization results for the states in $\epsilon_{if}$. 
As far as localization of eigenstates of chaotic systems are concerned, it is known 
that either all the states are exponentially localized (as in KR) or power-law 
localized (as in systems with singular potentials) but, to the best of our
knowledge, combinations of these localization profiles have not been reported before.
In the next section, we transform the Hamiltonian in Eq. \ref{ham1} to that of a
tight binding model and show that $V_0/E$ controls the localization property of eigenvectors.

\section{Tight binding model}			
The dynamical localization in the quantum KR system was mapped to 
Anderson model for electron transport in a one-dimensional crystalline lattice \cite{fishman}.
By implication, the exponential decay profile of eigenstates in the Anderson model 
translates to exponential profile (in the momentum representation) for the 
Floquet states of quantum KR.
Following this mapping technique, in this section, we map the Hamiltonian in Eq. \ref{ham1}
to a tight binding Hamiltonian.
Since Eq. \ref{ham1} represents a time periodic system, using Floquet-Bloch 
theorem, we can write the quasi-energy state as
\begin{equation}
\phi(\theta,t)=e^{-i\omega t} u(\theta,t),
\label{eq1}
\end{equation}
where, $u(\theta, t) = u(\theta, t+1)$. In between two consecutive
kicks, the Hamiltonian $H_0$ governs the evolution of the particle and is given by,
\begin{equation}
\phi_n^{-}(t+1)=e^{-iE_n} \phi_n^{+}(t).
\label{eq2}
\end{equation}
In this, $\phi_n^{-}(t+1)$ and $\phi_n^{+}(t)$ are the quasi-energy states just
before the $(t+1)$-th kick and just after $t$-th kick and $E_n$ is the $n$-th energy
level of $H_0$. During the evolution it acquires an extra phase $e^{-iE_n}$.
By substituting Eq. \ref{eq1} in Eq. \ref{eq2} and using the periodicity of $u(\theta, t)$, 
we obtain
\begin{equation}
\begin{aligned}
u_n^{-}(\theta, t+1)=e^{i \omega} e^{-i E_n} u_n^{+}(\theta, t).
\end{aligned}
\label{eq3}
\end{equation} 
			
Now the quasi-energy state just after a $t$-th kick can be obtained using a map
$\phi^{+}(\theta, t) = e^{-iV(\theta)} \phi^{-}(\theta, t)$.
By using Eq. \ref{eq1}, this can be written in-terms of $u(\theta, t)$ as 
\begin{equation}
u^{+}(\theta, t)=e^{-iV(\theta)} u^{-}(\theta, t).
\label{eq4}
\end{equation}

Now, $e^{-iV(\theta)}$ is expressed in terms of trigonometric function
$W(\theta)=-\tan\left(\frac{V(\theta)}{2} \right)$ as
\begin{equation}
e^{-iV(\theta)}=\frac{1+iW(\theta)}{1-iW(\theta)}.
\label{eq5}
\end{equation}
This is used in Eq. \ref{eq4} to obtain
\begin{equation}
\frac{u^{+}(\theta)}{1+i W(\theta)}=\bar{u}=\frac{u^{-}(\theta)}{1-i W(\theta)},
\end{equation}
where $\bar{u}$ is defined as $\bar{u}=[u^{+}(\theta)+u^{-}(\theta)]/2$.
Using Eq. \ref{eq4} and Eq. \ref{eq5}, the evolution of the quasi-energy state after 
one period is
\begin{equation}
u^{+}(\theta)=e^{-iV(\theta)} e^{i(\omega -E_n)} u^{+}(\theta).
\end{equation}
This can be written as,
\begin{equation}
(1-i W(\theta)) \bar{u}=e^{i(\omega -H_0)} \bar{u} (1+i W(\theta)),
\label{and1}
\end{equation}
where $\bar{u}=\frac{u^{+}}{1+i W(\theta)}$. 
Now rearrangement of terms leads to
\begin{equation}
\tan\left(\frac{\omega - H_0}{2} \right) \bar{u} + W(\theta) \bar{u} = 0.
\label{and2}
\end{equation}

The quasi-energy state can be expanded in the unperturbed basis as $|\bar{u}\rangle=\sum_m u_m |\psi_m\rangle$
where, $|\psi_m\rangle$ are the eigenstates of $H_0$ and $u_m$ is given by,
\begin{equation}
u_m = \int \bar{u} \psi_m(\theta) d\theta = \int \frac{1}{2} [u^{+}(\theta) + u^{-}(\theta)] \psi_m(\theta)~d\theta.
\end{equation}
Taking the inner product of Eq. \ref{and2} with $|\psi_m\rangle$, we will formally obtain
\begin{equation}
T_m u_m + \sum_l W_{ml} u_l = 0.
\label{tight1}
\end{equation}
In this, $T_m=\tan(\frac{\omega - E_m}{2})$ represents the on-site energy and $W_{ml}$ is the
hopping strength for a particle to hop from $m$th site to $l$th site and can be written in the
energy basis as
\begin{equation}
\begin{aligned}
W_{ml} &= \langle \psi_m| W(\theta) | \psi_l \rangle \\
&= \int \sum_{p,q} a^{*}_{mp} e^{-ip\theta} W(\theta) a_{lq} e^{i q \theta} d\theta \\
&= \sum_{p,q} a_{mp}^{*} a_{lq} \int W(\theta) e^{-i(p-q) \theta} d\theta \\
&= \sum_{p,q} a_{mp}^{*} a_{lq} W_{p-q},
\end{aligned}
\end{equation}
where $W_n=\frac{1}{2 \pi}\int_0^{2 \pi} W(\theta) e^{-i n \theta} d\theta$ is the
Fourier transform of $W(\theta)$.
Thus, in energy basis, after simple manipulation, Eq. \ref{tight1} takes the form
\begin{equation}
\left(T_m + \sum_{p,q} a_{mp}^{*} a_{mq} W_{p-q} \right) u_m + \sum_{p,q,l\neq m} a_{mp}^{*} a_{lq} W_{p-q} u_l = 0.
\label{tight2}
\end{equation}
This is the tight binding model version of the Hamiltonian in Eq. \ref{ham1}. In this, 
$(T_m + \sum_{p,q} a_{mp}^{*} a_{mq} W_{p-q})$ represents the diagonal term and 
$a_{mp}^{*} a_{lq} W_{p-q}$ is the off-diagonal term of the transfer matrix. 
It does not appear straightforward to analytically prove power-law profile of
Floquet states
starting from Eq. \ref{tight2}, though it appears fair to expect that
in this case the decay of Floquet state profile will be different from exponential form. As numerical results
show, we obtain power-law localization. Similar results has also been reported in \cite{cohen}.

However, using Eq. \ref{tight2}, it is possible to make an inference about Floquet state profile
in the limit $\mu >> 1$.
In this case $E_n \gg V_0$ and the singularity in the potential becomes
insignificant.  Effectively, the system behaves as a free particle with energy
$E_n=\frac{\hbar_s n^2}{2}$ and the wave-function of $H_0$ is just the momentum eigenstate,
$|\psi_n\rangle = a_{nn} e^{i n \theta}$, with $a_{mn}=\delta_{mn}$.
This set of conditions, if applied to Eq. \ref{tight2}, lead to 
\begin{equation}
(T_m + W_0 ) u_m + \sum_l W_{m-l} u_l = 0.
\end{equation}
This is just standard KR Hamiltonian transformed to the 1D Anderson model \cite{fishman}, for which
all the eigenstates are known to display exponential profile. Hence,
as seen in Fig. \ref{avgdrm}(c) for $\mu \gg 1$, the observed localization is exponential in nature.
Thus, even in the presence of singular potentials, eigenstate localization is not
generically of power-law form. We reiterate the main result of the paper that the  association
between power-law profile of eigenstates and singular potentials needs to take into account
the effectiveness of singularity in a given energy band.
			
\section{Spectral signatures}

  Based on the results presented in Fig. \ref{avgdrm}, a novel scenario for the
spectral signatures can be expected. As the regimes $\mu < 1$ and $\mu \gg 1$ are
traversed, by considering Floquet states in a suitable energy band $\epsilon_{if}$, the decay profile 
of Floquet state changes from power-law to exponential form. This would also imply that 
a unique spectral signature for the nearest neighbour spacing distribution $P(s)$, such 
as either the Poisson or Wigner distributions, may not exist for the system as a whole. Quite
unusually, $P(s)$ would depend on the energy band $\epsilon_{if}$ being considered.
Thus, in the same system for a given choice of parameters, in the limit $\mu \gg 1$ (KR limit)
we expect Poisson distribution and in the limit $\mu < 1$ (KRIW limit) we expect P(s) to be 
closer to Wigner distribution or possibly, a Brody distribution \cite{brody}. 

The Floquet operator $\widehat{U}$ being a unitary operator, all the
eigenvalues lie on a unit circle, $\omega_i\in[0,2\pi)$. In this case, level
density is constant $\left(\frac{N}{2\pi} \right)$ and hence the unfolding of Floquet
levels is not necessary. To compute the spacing distribution, we have treated the 
eigenvalues of even and odd parity states separately. The nearest neighbour spacing 
distribution reveals two
different forms; for $\mu < 1$ level repulsion is observed in the form of Brody distribution
and for $\mu \gg 1$ level clustering is seen in the form of Poisson distribution \cite{supplement}.
The regime of $\mu < 1$ corresponds to KRIW limit and power-law decay of Floquet
states (see Fig. \ref{avgdrm}(b)) and is associated with level correlations that
are intermediate between no correlation and random matrix type level repulsion.
On the other hand, the limit of $\mu \gg 1$ is KR limit and levels remain uncorrelated
due to occurrence of dynamical localization.
It must be emphasised that two different level spacing distributions and level correlations 
for the same system with identical parameters is a novel feature not usually
encountered in the context of chaotic quantum systems. This unusual spacing distribution
reinforces the central result of this paper that the relation between potential singularity 
and eigenvector profile is conditioned by energy regime being considered.

\begin{figure}
\centering
\includegraphics*[width=3.0in]{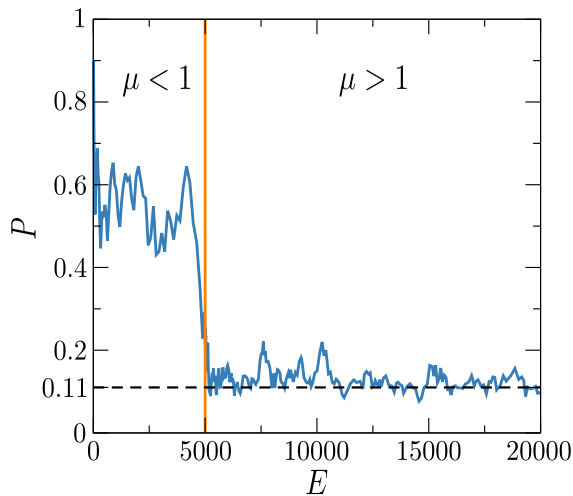}
\caption{(Color online) Participation ratio of the Floquet states as a function of $E_{max}$ 
for $b=1.4\pi$, $\hbar_s = 1.0$, $V_0=5000.0$, $k=4.25$ (same set of parameters
as in Fig. \ref{avgdrm}). The vertical line is placed at $E_{max}=V_0$ ($\mu=1$), the height 
of potential barriers. Horizontal black dotted line represents the mean participation ratio for $\mu>1$.}
\label{PR}
\end{figure}

This dichotomy is reflected in the eigenvector statistics as well.
This is easily observed by studying the participation ratio (PR) of the Floquet states
that provides information about their localization properties. For an eigenstate that resides
in the infinite dimensional Hilbert space, participation ratio is defined as
\begin{equation}
P=\sum_{i=1}^{\infty} |\psi_i|^4
\end{equation}
with the condition that $\sum_i |\psi_i|^2 = 1.0$, where $\psi_i$ are components 
of a Floquet state. It is a measure of how many basis states
effectively participate in making up the eigenstate. If $P \approx 1$, then
the state is strongly localized and implies that one basis state contributes significantly 
to the Floquet state
while the contribution from the rest of the basis are almost negligible. However, if 
$P \sim \frac{1}{N}$, then the Floquet state is of extended nature and all the 
basis states make equal contribution on an average.
Figure \ref{PR} displays $P$ for all the $10035$ converged Floquet states as a function
of energy $E_{max}$ for the identical choice of parameters as in Fig. \ref{avgdrm}.
Quite surprisingly, $P$ distinguishes the two regimes, $\mu < 1$ and $\mu \gg 1$. The
boundary between the two regimes is at $E_{max}=V_0$, the height of potential barriers.
For $\mu \gg 1$, exponential localization of Floquet states implies that
$|\phi\rangle \sim e^{-n/l}$, where $l$ is the localization length. A remarkable result due
to Izrailev \cite{kr, chirikov} provides the relation, $l \approx \frac{k^2}{2 \hbar_s^2}$. For our case, this estimate
gives $l \approx \frac{k^2}{2 \hbar_s^2} = 9.03$
and this represents the effective number of basis states that goes in constructing
the Floquet states. As participation ratio is the inverse of the effective number
of basis states, it is estimated to be $P \approx \frac{2 \hbar_s^2}{k^2} = 0.11$. As seen in Fig. \ref{PR},
this value closely matches the computed PR in the regime $\mu \gg 1$.

For $\mu<1$, the mean $P$ is larger compared to that for $\mu>1$ as shown in Fig. \ref{PR}.
The reason can be traced back to the fact that in the case of infinite well $E_n \sim n^2$
and hence levels are spaced far apart.
This implies that the Floquet states for $\mu<1$ has overlap only with a few unperturbed
basis states and this effectively increases the value of participation ratio for $\mu<1$. 
Ultimately, this results in a more compact localization.

\begin{figure}
\centering
\includegraphics*[width=3.0in]{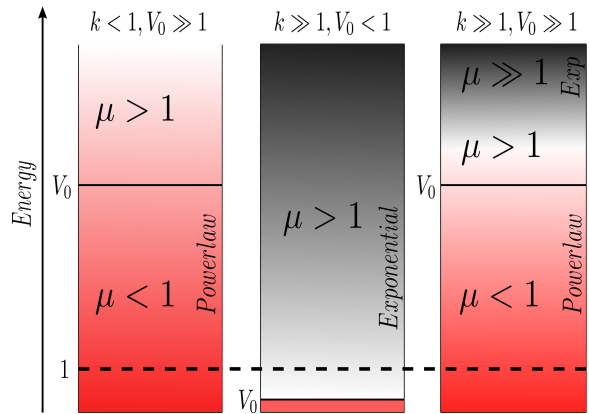}
\caption{(Color online) Summary of the results presented in this work. In all the cases,
the potential $V_{sq}$ is singular. Power-law decay profile
of the Floquet is obtained whenever $\mu < 1$ (shown as deep red). Stronger red color
represents power-law profile over larger energy scales. For $\mu > 1$, exponential localization
is obtained. Darker grey represent dominant exponential profile over longer energy scales. 
White color represents regimes of transition between these profiles. They cannot be classified
as power-law or exponential with definiteness. Black broken horizontal line corresponds to $V_0=1$} 
\label{sumfig}
\end{figure}

Finally, all the results discussed in this paper can be summarised in the form of a 
`phase diagram' displayed in Fig. \ref{sumfig}. For $\mu <1$, singularity in the
potential is effective and hence power law profile of the Floquet states is obtained.
This regime is indicated by red color in Fig. \ref{sumfig}.
However, if $\mu > 1$, singularity is not effectively `felt' by the particle and
hence exponential profile is obtained. This regime is indicated by black color 
in the figure. Depending on the choice of parameters, regimes
in which transition occurs between these two Floquet state profiles are also
observed. In Fig. \ref{sumfig}, this is indicated by white color.

\section{Conclusion}
In summary, we have studied a non-KAM system represented by the Hamiltonian in Eq. \ref{ham1}, 
namely a periodically kicked particle in a finite potential well of height $V_0$, to primarily understand 
the nature of its Floquet states.  This Hamiltonian can be thought of as representing two limiting
cases, (i) the standard KR for $V_0=0$ and (ii) KR in infinite potential
well for $V_0 \to \infty$. It is well known that, for sufficiently large kick strengths, all the 
Floquet states of the KR are localized with an exponential profile \cite{kr}. Further, it has been
suggested that for kicked systems with singularity in their potential, the Floquet states display
power-law profile \cite{liu}. We examine the Floquet states of the Hamiltonian in Eq. \ref{ham1} in the light
of these results. To understand its Floquet states, we map this problem to that of a tight binding
model. 

The results presented in this work show that the decay profile of the Floquet states
is not determined by the potential singularity alone, but by the representative energy band 
$\epsilon_{if}$ of a set of Floquet states relative to the potential height $V_0$. 
Thus, we show that if $\epsilon_{if} > V_0$ then the effect of singularity is weak for those
set of Floquet states and they display exponential profile. This represents the KR
limit of the problem. On the other hand, the condition $V_0 > \epsilon_{if}$ represents
Floquet states strongly affected by the singular potential. In this case, we have shown that 
Floquet states have predominantly power-law profile. In the region intermediate between these
two extremes, the Floquet states typically display an initial exponential decay followed by
an asymptotic power-law decay. The presence of these two contrasting Floquet state profiles
in Hamiltonian in Eq. \ref{ham1} leaves its signature in the spectral correlations as well.
For identically same set of parameters, depending on the reference energy scale $\epsilon_{if}$,
the spacing distribution turns out to be Poisson distribution ($\epsilon_{if} > V_0$) or a 
general Brody distribution ($V_0 > \epsilon_{if}$). Typically, the spacing distribution 
is taken to characterize quantum chaos in a system and it is generally independent of the energy
band being considered provided it is in the semi-classical limit.
Quite surprisingly, the semi-classical limit of the system in Eq. \ref{ham1} lacks a unique 
spacing distribution as it depends on the energy band $\epsilon_{if}$ being considered.
KR was experimentally realized in a test-bed of cold atomic cloud in flashing 
optical lattices. Using more than one optical lattice, KR confined to a `potential well'
has also been realized. We believe that the results in this work is amenable to experiments
in a suitable atom-optics set up.

\section{Acknowledgement}
S. P. would like to acknowledge the University Grants Commission for research fellowship.

\onecolumngrid

\newpage

\begin{center}
{\Large \bf \underline{Supplemental Material}}
\end{center}

		\section*{A. Stroboscopic section}
		In the main paper, we have considered a particle in a periodic potential well which is receiving periodic kicks. This is a non-KAM system. We mainly looked at the localization properties of this Floquet system. We have found that depending on the energy range considered, the eigenstates of the system show either power-law or exponential which is an unusual feature observed. We have also ensured that these systems are classically chaotic. Fig. \ref{ps} shows the stroboscopic section for the systems presented in Fig. 2 and Fig. 4 in the main paper. We can see that for all the three cases shown in Fig. \ref{ps}, the systems are classically chaotic. The phase space has been calculated using the map described in Ref. \cite{paul}. We have applied periodic boundary conditions at the positions A and B mentioned in Fig. 1 in the main paper.

		\begin{figure}[h]
		\begin{center}
		\includegraphics[width=6in]{fig7.eps}
		\caption{Stroboscopic sections of kicked particles in a periodic square well potential for $b=1.4\pi$, a) $V_0=5000.0$, $k=0.25$, b) $V_0=0.5$, $k=4.25$, and c) $V_0=5000.0$, $k=4.25$.}
		\label{ps}
		\end{center}
		\end{figure}

		\section*{B. Spacing distribution}

		\begin{figure}[h]
		\begin{center}
		\includegraphics[width=6in]{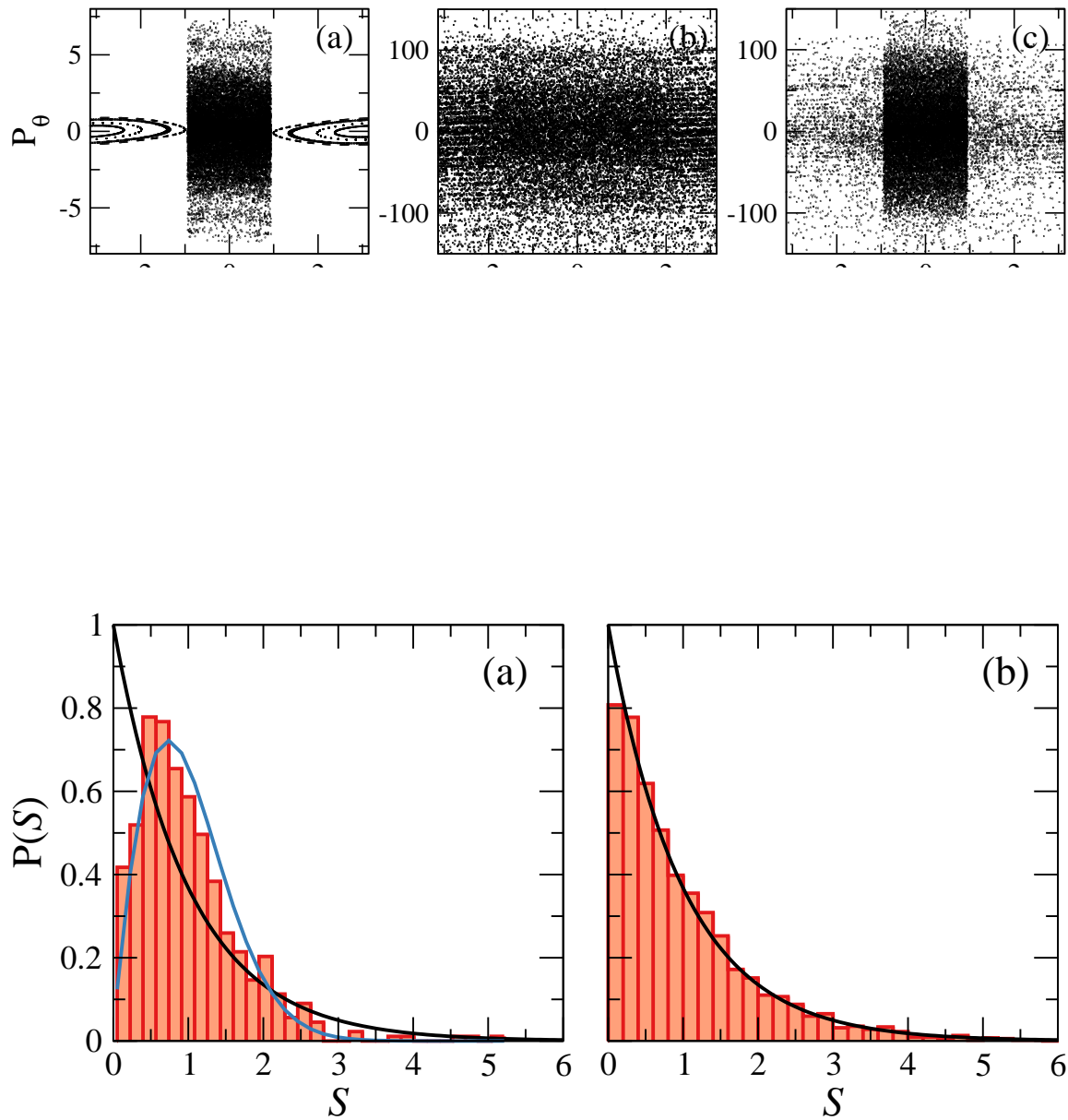}
		\caption{Spacing distribution $P(S)$ for $b=1.4\pi$, $k=4.25$, $V_0=1500000.0$, a) $\mu < 1$, b) $\mu \gg 1$. Black curve represents Poisson distribution and blue curve is the Brody fit.}
		\label{spacing}
		\end{center}
		\end{figure}

		In Fig. \ref{spacing} we have shown spacing distribution for $V_0=1500000$, $k=4.25$, $b=1.4\pi$ and $\hbar_s=1.0$. We have used a large value of $V_0$ in order to obtain a sufficient number of Floquet eigenvalues in the $\mu_1$ regime (mentioned in the main text) to calculate the spacing distribution. We have separated the even and odd parity states. Fig. \ref{spacing}(a) demonstrates that spacing distribution shows clear deviation from Poisson distribution ($P(S)=\exp(-S)$) which is represented by a black curve. We have fitted the Brody distribution \cite{reichl} (shown as blue curve) 
		
		\begin{equation}
		P(S)=A\left(\frac{S}{D}\right)^\alpha \exp\left(-a\left(\frac{S}{D}\right) ^{(1+\alpha)}\right)
		\end{equation}
					
		where $\alpha$ is the Brody parameter, $A=a(1+\alpha)$ and 
		$a=[\Gamma(2+\alpha)/(2+\alpha)]^{1+\alpha}$. The best fit value of Brody parameter is $\alpha=0.83$. This large value of $\alpha$ implies eigenvalues are no longer fully uncorrelated and hence show level repulsion. On the other hand, the spacing distribution in the $\mu_2$ regime (mentioned in the main text) follows Poisson distribution which implies eigenvalues are closely spaced or are uncorrelated. The eigenvectors are well separated. This situation resembles that of standard KR system where we obtain spacing distribution which follows Poisson distribution. Thus it is a clear evidence that the system at higher energy ($E\gg V_0$) behaves similarly to the kicked rotor system.

\end{document}